%% file: acs-latex-template.tex
\newif\ifpreprint
\preprintfalse 

\ifpreprint
\documentclass[journal=jpclcd,manuscript=letter]{achemso}
\else
\documentclass[journal=jpclcd,manuscript=letter,layout=twocolumn]{achemso}
\fi

\usepackage[T1]{fontenc} 

\usepackage{amsmath}
\usepackage{newtxtext,newtxmath}

\usepackage{pifont}
\usepackage{graphicx}
\usepackage{dcolumn}
\usepackage{braket}
\usepackage{multirow}
\usepackage{threeparttable}
\usepackage{xspace}
\usepackage{verbatim}
\usepackage[version=4]{mhchem} 
\usepackage{comment}
\usepackage{color,soul}

\usepackage{mathtools}
\usepackage[dvipsnames]{xcolor}
\usepackage{xspace}
\usepackage{ifthen}

\usepackage{qcircuit}

\usepackage{graphicx,longtable,dcolumn,mhchem}
\usepackage{rotating,color}
\usepackage{lscape}
\usepackage{amsmath}
\usepackage{dsfont}
\usepackage{soul}
\usepackage{physics}

\newcommand{\SI}{\textcolor{blue}{Supporting Information }}

\def\nonum{\nonumber \\}

\def\beit{\begin{itemize}}
\def\eit{\end{itemize}}

\def\mbb{\mathbb}

\def\tb{\textcolor{black}}

\usepackage[colorlinks = true,
            linkcolor = blue,
            urlcolor  = black,
            citecolor = blue,
            anchorcolor = black]{hyperref}
\definecolor{goodorange}{RGB}{225,125,0}
\definecolor{goodgreen}{RGB}{5,130,5}
\definecolor{goodred}{RGB}{220,50,25}
\definecolor{goodblue}{RGB}{30,144,255}

\newcommand{\note}[2]{
\ifthenelse{\equal{#1}{F}}{
\colorbox{goodorange}{\textcolor{white}{\footnotesize \fontfamily{phv}\selectfont #1}}
    \textcolor{goodorange}{{\footnotesize \fontfamily{phv}\selectfont #2}}\xspace
}{}
\ifthenelse{\equal{#1}{R}}{
\colorbox{goodred}{\textcolor{white}{\footnotesize \fontfamily{phv}\selectfont #1}}
    \textcolor{goodred}{{\footnotesize \fontfamily{phv}\selectfont #2}}\xspace
}{}
\ifthenelse{\equal{#1}{N}}{
\colorbox{goodgreen}{\textcolor{white}{\footnotesize \fontfamily{phv}\selectfont #1}}
    \textcolor{goodgreen}{{\footnotesize \fontfamily{phv}\selectfont #2}}\xspace
}{}
\ifthenelse{\equal{#1}{M}}{
\colorbox{goodblue}{\textcolor{white}{\footnotesize \fontfamily{phv}\selectfont #1}}
    \textcolor{goodblue}{{\footnotesize \fontfamily{phv}\selectfont #2}}\xspace
}{}
}

\usepackage{titlesec}

\usepackage[fontsize=11pt]{scrextend}
\titleformat{\section}
{\normalfont\sffamily\bfseries\color{Blue}}
{\thesection.}{0.25em}{\uppercase}

\titleformat{\subsection}[runin]
{\normalfont\sffamily\bfseries}
{\thesubsection}{0.25em}{}[.\;\;]

\titleformat{\suppinfo}
{\normalfont\sffamily\bfseries}
{\thesubsection}{0.25em}{}

\titlespacing*{\section}{0pt}{0.5\baselineskip}{0.01\baselineskip}
\titlespacing*{\subsection}{0pt}{0.125\baselineskip}{0.01\baselineskip}

\author{Luis Martinez-Gomez}
	\affiliation{Department of Chemistry and Cherry Emerson Center for Scientific Computation, Emory University, Atlanta, GA, 30322, USA}
\author{Raphael F. Ribeiro}
    \email{raphael.ribeiro@emory.edu}
	\affiliation{Department of Chemistry and Cherry Emerson Center for Scientific Computation, Emory University, Atlanta, GA, 30322, USA}

\let\oldmaketitle\maketitle
\let\maketitle\relax
	\title{Topological advantage for adsorbate chemisorption on conjugated chains}
\date{\today}

\begin{tocentry}
	\vspace{1cm}
	\includegraphics[width=8cm]{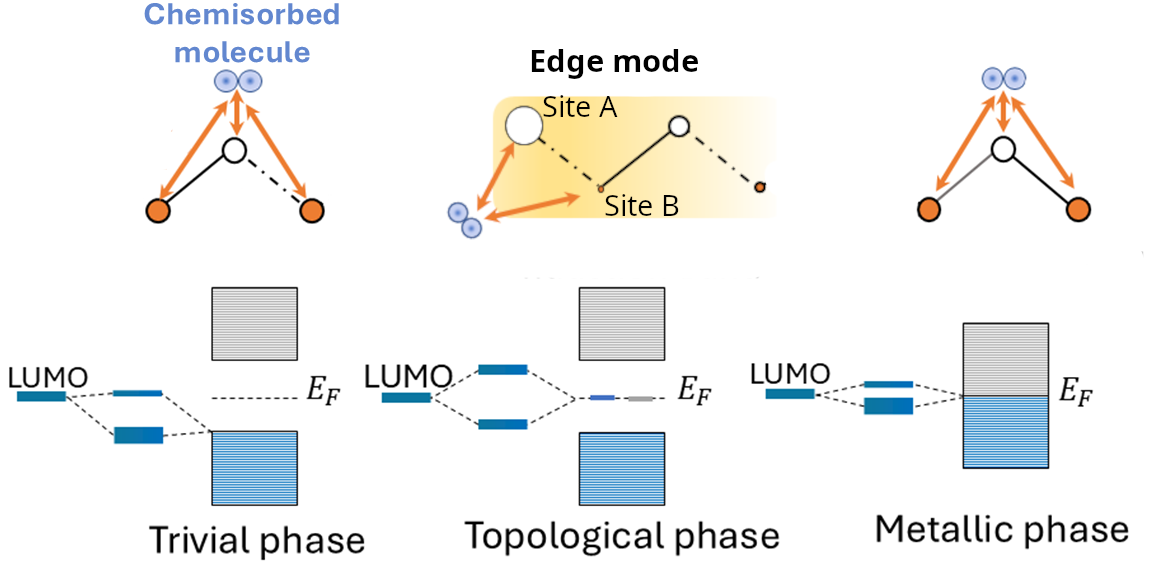}
\end{tocentry}

\begin{document}	

\ifpreprint
\else
\twocolumn[
\begin{@twocolumnfalse}
\fi
\oldmaketitle


\begin{abstract}
Topological matter offers opportunities for control of charge and energy flow with implications for chemistry still incompletely understood. In this work, we study an ensemble of adsorbates with an empty frontier level (LUMO) coupled to the edges, domain walls (solitons), and bulk of a Su–Schrieffer–Heeger polyacetylene chain across its trivial insulator, metallic, and topological insulator phases. We find that two experimentally relevant observables, charge donation into the LUMO and the magnitude of adsorbate electronic friction, are significantly impacted by the electronic phase of the SSH chain and show clear signatures of the topological phase transition. Localized, symmetry-protected midgap states at edges and solitons strongly enhance electron donation relative to both the metallic and trivial phases, whereas by contrast, the metal’s extended states, despite larger total DOS near the Fermi energy, hybridize more weakly with a molecular adsorbate near a particular site. Electronic friction is largest in the metal, strongly suppressed in gapped regions, and intermediate at topological edges where hybridization splits the midgap resonance. These trends persist with disorder highlighting their robustness and suggest engineering domain walls and topological boundaries as pathways for employing topological matter in molecular catalysis and sensing.

\end{abstract}

\ifpreprint
\else
\end{@twocolumnfalse}
]
\fi

\ifpreprint
\else
\small
\fi

\noindent


\input{introduction}


\input{theory}


\input{results_and_discussion}


\input{conclusions}

\textbf{Acknowledgments}
R.F.R. acknowledges generous start-up funds from Emory University.

\begin{suppinfo}
\tb{We provide evidence of lack of size dependence of our results by reporting LUMO electronic occupation numbers  and electronic friction versus SSH chain length $N$ and discuss the Gaussian broadening $\sigma$ used in the friction calculation, including its effect on the adsorbate-projected DOS in the trivial and topological phases.}
\end{suppinfo}

\bibliography{References.bib}

\end{document}

%% file: introduction.tex
Topological insulators are electronic phases of matter with a bulk band gap and gapless surface states protected by global symmetries \cite{hasan2010colloquium, moore2010birth, schnyder2008classification, asboth2016short, kane2005quantum, kane2005z, bernevig2006quantum}. These materials exhibit intriguing properties, including high surface carrier mobility \cite{shekhar2012ultrahigh, shekhar2012electronic, analytis2010two, thio1998giant, butch2010strong, jia2011low, li2012electron} with low power dissipation \cite{roth2009nonlocal, chang2015zero}, and spin-polarized currents with unconventional textures \cite{ando2014electrical, tian2015electrical, souma2011direct, pan2011electronic}. Potential applications include spintronics \cite{hsieh2009observation, he2019topological, fan2016spintronics, he2022topological, yokoyama2014spintronics}, quantum computing \cite{he2019topological, paudel2013three, scappucci2021crystalline}, and thermoelectrics \cite{xu2017topological, ivanov2018thermoelectric, matsushita2017thermoelectric, muchler2013topological, xu2016thermoelectric}. Recently, there has also been interest in utilizing topological insulators and semimetals in electrochemistry and chemical catalysis \cite{rajamathi2017weyl, schoop2018chemical, li2020heterogeneous, muechler2020topological, narang2021topology, zhang2023topological, weng2024understanding, li2023topological, li2024synergistic, lau2025facet, laderer2025topological}. Their highly localized conducting boundary modes may significantly influence surface reactions, and their properties are protected against local perturbations (e.g., lattice defects and impurities). For photocatalysis, high surface mobility is desirable because efficient charge separation and suppressed electron–hole recombination can enhance performance \cite{zhang2016mnpse3, lv2017two, li2020heterogeneous, rouzhahong2020first}. Indeed, experimental \cite{chen2013study, rajamathi2017weyl, li2019surface, he2019topological1, yang2019electrochemical} and computational \cite{chen2011co, Li2017, weng2024understanding} studies provide evidence that, under certain conditions, topological matter exhibits enhanced reactivity mediated by topologically protected boundary electrons. Nevertheless, the extent to which the surfaces of topological materials can be systematically exploited for high-efficiency, selective synthesis remains an open question.

In heterogeneous catalysis, adsorbed molecules undergo several processes on a solid surface \cite{hoffmann1988chemical, ertl1990elementary, jiang2019dynamics, ertl1997handbook}. A central step is charge transfer between the extended material and the adsorbates \cite{weisz1953effects, lewis2005chemical, wodtke2008energy}. Orbital hybridization and electron donation into a molecular system influence adsorption \cite{sung1985carbon, hoffmann1988chemical, santen1991theoretical, jung2021understanding}, surface diffusion \cite{mckenna2010interplay, wahlstrom2004electron}, and desorption, and can reduce energetic barriers for bond breaking or association \cite{kasemo1996charge, dai1995laser}. Likewise, adsorbate–surface energy exchange via vibrational relaxation and excitation of boundary electron–hole pairs (EHPs) \cite{newns1969self, bohnen1975friction, persson1980vibrational, hellsing1984electronic, tully1990molecular, tully1995molecular, tully2000chemical, alec2004electronically, dou2020nonadiabatic, ghan2023interpreting, meng2025dynamics, preston2025nonadiabatic} has been implicated in $\text{H}_2$ relaxation on metal surfaces \cite{persson1982electronic, juaristi2008role}, $\text{CO}_2$ reduction \cite{nam2020molecular, schneider2012thermodynamics, diercks2018role, ge2018electron}, and water oxidation \cite{materna2017anchoring}. Here we focus on two mechanisms that are important to heterogeneous catalysis and can be directly analyzed in a minimal topological setting: (i) charge hybridization between molecular frontier orbitals and boundary electronic states, and (ii) nonadiabatic vibrational energy dissipation (electronic friction) mediated by EHPs. We ask how these processes are modified when the substrate hosts symmetry-protected boundary modes versus trivial or metallic electronic structures.

\tb{We employ the Su--Schrieffer--Heeger (SSH) model \cite{su1979solitons, heeger1988solitons} as a minimal platform in which topological, metallic, and trivial insulating regimes arise within a single Hamiltonian, while still allowing controlled introduction of defects and systematic thermodynamic-limit analysis. Although the SSH chain is one-dimensional and we use a spinless, effective single-particle form, it faithfully captures the chiral (sublattice) symmetry and the associated boundary- and domain-wall-localized midgap orbitals that constitute the central ingredient of this work and that, as we show below, strongly modulate molecule--substrate interfacial properties.}

\tb{This effective description is particularly well motivated for light-element $\pi$-conjugated polymers such as trans-polyacetylene: intrinsic spin--orbit coupling in the carbon $p_z$ manifold is weak compared with eV-scale hoppings \cite{boehme2013challenges}, and electron--electron interactions \cite{bendazzoli1999density} can be treated at an effective level through renormalized hopping amplitudes and onsite energies within a tight-binding picture. We briefly comment at the end on extensions to spinful models with strong spin--orbit coupling and/or strong correlations. The resulting minimal-model approach is therefore designed to isolate robust, design-relevant trends, and in particular to clarify when and where enhanced hybridization and nonadiabatic dissipation should be expected, in a way that enables a direct interpretation of the effects reported here.}

\tb{Within this controlled setting, we couple molecular adsorbates to SSH-class substrates to isolate and quantify how symmetry-protected midgap states (at finite edges and at topological domain walls) govern molecule-substrate interfacial observables. Across metallic, trivial insulating, and topological regimes, and in the presence of both trivial and topological defects, we compute adsorbate level broadening (hybridization), LUMO occupancy, and charge transfer for single and multiple adsorbates, thereby disentangling the roles of the density of states and localized midgap modes. We further evaluate the electronic friction acting on an adsorbate within linear response, via a kernel that captures electron--hole-pair--mediated vibrational damping at the contact. We show that localized midgap modes can (i) strongly enhance adsorbate hybridization and charge donation relative to both trivial insulating and metallic substrates, and (ii) sharply increase vibrational energy dissipation through enhanced electronic friction upon crossing the topological transition; moreover, topological domain walls mitigate the potential measure-zero limitation of edge modes by providing a potentially extensive, chemically accessible set of such midgap states.}


%% file: theory.tex
\begin{center}
    \begin{figure}[hbt]
    \includegraphics[width=8.5cm]{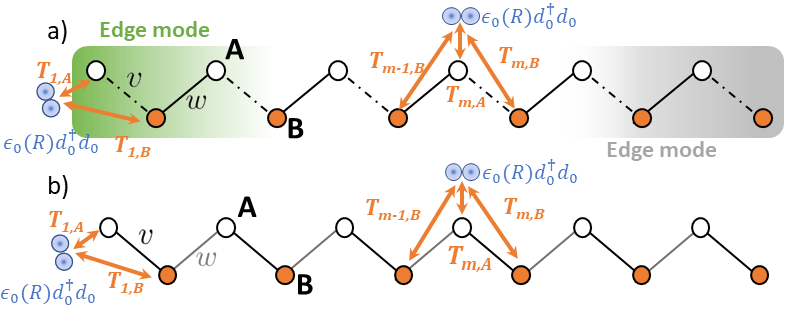}
    \caption{Schematic Fano--Anderson SSH setup. White and orange circles denote sublattices A and B, respectively, each with a single $p_z$ orbital. Intra- and intercell hoppings ($v$ and $w$) are tunable. A representative diatomic adsorbate couples either near an edge ($x_M=x_E$) or at the chain center ($x_M=x_B$). (a) Hybridization with edge/bulk modes in the topological phase ($v<w$). (b) Hybridization with a trivial insulator ($v> w$) or with the metallic limit ($v=w$), where states are delocalized.}
    \label{fig:Geometry}
    \end{figure}
\end{center}

\noindent\textbf{Fano-Anderson SSH model.} \label{sec:theory}
The simplest scenario we considered consists of a molecule adsorbed at different regions of an open conjugated SSH chain (Fig.~\ref{fig:Geometry}). The isolated molecule has a closed shell and a single low-lying empty electronic orbital, while the SSH chain models the extended material \cite{su1979solitons, su1980soliton}. The molecule interacts with nearby lattice sites as in the Fano--Anderson model \cite{fano1961effects, anderson1961localized}. To probe the roles of bulk versus boundary modes, we place the molecule either near the chain edge or at the center (Fig.~\ref{fig:Geometry}).

The SSH model represents a polyacetylene chain by a 1D tight-binding lattice with two identical sites per unit cell, nearest-neighbor staggered hopping \cite{su1979solitons}, and one electron per site. For a spinless open system, the isolated chain Hamiltonian is
\begin{align} \label{eq:hamiltonian_ssh}
    \hat{H}_{\mathrm{SSH}} &= v\sum_{j=1}^N\!\left( \hat{d}_{j,B}^\dagger \hat{d}_{j,A}+ \hat{d}_{j,A}^\dagger\hat{d}_{j,B} \right) \nonumber \\
     &\quad + w\sum_{j=1}^{N-1}\!\left( \hat{d}_{j+1,A}^\dagger \hat{d}_{j,B}+ \hat{d}_{j,B}^\dagger \hat{d}_{j+1,A} \right),
\end{align}
where $v$ ($w$) is the intra\-cell (intercell) hopping, $N$ is the number of unit cells, and $\hat{d}_{j,\alpha}^\dagger$ ($\hat{d}_{j,\alpha}$) creates (annihilates) an electron on sublattice $\alpha\in\{A,B\}$ in cell $j$. The gapped phases correspond to $v\neq w$: $v>w$ is a trivial insulator without boundary-localized modes, whereas $v<w$ yields a gapped system with two zero-energy boundary modes \cite{su1979solitons, asboth2016short}. 

\tb{In a fully spinful description of a dimerized $\pi$-conjugated chain (e.g., polyacetylene), each spatially localized midgap bound state corresponds to a localized $\pi$ frontier orbital. Depending on filling and interactions, this orbital can be singly occupied and carry an unpaired electron, i.e., it acts as a radical center in the language of conventional organic chemistry \cite{bredas1984role}. In the present work, we use a spinless SSH Hamiltonian, which can be viewed as isolating a single spin channel. In a more realistic spinful model the same spatial orbital would be twofold spin-degenerate. Hence, the topological midgap states of the SSH chain are tight-binding analogues of chemically familiar radical-like frontier orbitals, with the dimerization ratio ($w>v$ vs.\ $w<v$) controlling whether such orbitals are present or absent. In what follows, we focus on how the existence, localization, and filling of these midgap frontier orbitals modulate charge transfer and electronic friction between the conjugated backbone and the adsorbate.}

The isolated effective single-level adsorbate is described by $\hat{H}_0(R)=\varepsilon_0(R)\,\hat{d}_0^\dagger \hat{d}_0$, where $\hat{d}_0$ annihilates an electron in the adsorbate (LUMO) and $R$ is an internal vibrational coordinate. The full  Hamiltonian for a single species adsorbed to the SSH chain is
\begin{equation} \label{eq:FA_Hm}
    \hat{H}(R, x_M) = \hat{H}_{\mathrm{SSH}} + \hat{H}_0(R) + \hat{W}(x_M),
\end{equation}
where $x_M\in\{x_B,x_E\}$ labels bulk- or edge-coupled configurations. The interaction Hamiltonian $\hat{W}(x_M)$ describes hopping between the chain and the adsorbate orbital:
\begin{align}
    \hat{W}(x_E)  &=  T_{1,A}\,\hat{d}_{1,A}^\dagger \hat{d}_0 \;+\; T_{1,B}\,\hat{d}_{1,B}^\dagger \hat{d}_0 \;+\; \text{h.c.}, \\
    \hat{W}(x_B) &= T_{N/2,A}\,\hat{d}_{N/2,A}^\dagger \hat{d}_0 \;+\; T_{N/2,B}\,\hat{d}_{N/2,B}^\dagger \hat{d}_0 \nonumber\\
    &\quad + T_{N/2-1,B}\,\hat{d}_{N/2-1,B}^\dagger \hat{d}_0 \;+\; \text{h.c.},
\end{align}
where $N$ is taken even for simplicity and the couplings $T_{s,\alpha} \in \mbb{R}$ are positive and much smaller than $v$ and $w$. Figure~\ref{fig:Geometry} summarizes the considered configurations.

Polyacetylene chains may host both trivial and topological lattice defects \cite{su1979solitons, heeger1988solitons}. The latter are domain walls separating trivial ($v>w$) and topological ($v<w$) regions and support mid-gap localized modes. We therefore also consider a scenario where multiple molecules interacting with such defects. In the considered static limit (immobile defects), $N_d$ solitons centered at sites $m_{1},\dots,m_{N_d}$ are modeled by site-dependent nearest-neighbor hoppings 
\begin{align}
    t_n &\equiv H _{\mathrm{SSH}}^{\,n,n+1} 
    = \frac{w+v}{2} + \frac{v-w}{2}\sum_{i=1}^{N_d}\phi_n^{(m_i)},
\end{align}
with \cite{takayama1980continuum}
\begin{align}
    \phi_n^{(m_i)} = (-1)^n \tanh\!\left(\frac{n-m_i}{\xi}\right),  \label{eq:phinm0}
\end{align}
where $\xi$ is the soliton width. \tb{Although solitons in real polyacetylene generally arise as thermally activated defects, here we treat them as fixed static inhomogeneities and work at half filling with a fixed electron number. This approximation allows us to examine the fundamental role of topological defects without introducing additional assumptions regarding thermally driven soliton formation and dynamics.}

\par In disordered samples, each adsorbate is placed at a soliton center $m_i$ and couples to the nearest site with strength $T_{m_i}$ and to its two nearest neighbors $m_i\!\pm\!1$ with $T_{m_i}/3$, so each adsorbate interacts with three sites. The adsorbates are typically sufficiently far from each other that the adsorbate ensemble Hamiltonian is the the straightforward generalization of $\hat{H}_0$ to multiple noninteracting adsorbates.

\noindent\textbf{Observables.}
To quantify charge transfer and hybridization for a single adsorbate, we compute the adsorbate population $n_0(R,x_M)$ for a half-filled SSH chain at zero temperature with the Fermi energy set to $\mu=0$. Defining the occupied single-particle projector as $\hat{P}_{\mathrm{occ}}(R,x_M)=\Theta\!\big(\mu-\hat{H}(R,x_M)\big)$, where $\Theta$ is the Heaviside step function, we write
\begin{align}
    n_0(R,x_M) 
    &= \langle \varepsilon_0 \,|\, \hat{P}_{\mathrm{occ}}(R,x_M)\, |\, \varepsilon_0\rangle \nonum 
    &= \sum_{\lambda:\,\epsilon_\lambda(R,x_M)<\mu}\big|\langle \varepsilon_0 \,|\, \lambda(R,x_M)\rangle\big|^2,
\end{align}
where $|\varepsilon_0\rangle$ is the single-particle orbital associated with the adsorbate creation operator $\hat{d}_0^\dagger$, and $\{|\lambda(R,x_M)\rangle\}$ are eigenstates of $\hat{H}(R,x_M)$ with energies $\{\epsilon_\lambda(R,x_M)\}$. In practice, we take $\mu\rightarrow 0^-$ to resolve any zero-mode degeneracies in the topological phase. We obtain $n_0(R,x_M)$ by numerical diagonalization and, where appropriate, perturbation theory.

For multiple adsorbates and a variable number of defects, we assess chemisorption via the mean LUMO occupancy
\begin{align}
    \braket{n_0} = \frac{1}{N_{\mathrm{ad}}}\sum_{i=1}^{N_{\mathrm{ad}}} n_{0i},
    \label{eq:average_n0}
\end{align}
where $n_{0i}$ is the $i$th adsorbate LUMO population.

To probe vibrational energy exchange with the substrate, we consider the adsorbate vibrational degree of freedom $R$ as a dynamical variable, modeled as a harmonic oscillator with mass $m$, frequency $\omega$, and displacement from equilibrium (at the electronic ground-state) $R$. We assume a linear vibronic coupling of the adsorbate level \cite{persson1982electronic, dou2020nonadiabatic},
\begin{align}\label{Eq:epsilon0R}
& \hat{H}_0(R) = \varepsilon_0(R)\,\hat{d}_0^\dagger\hat{d}_0 + U_0(R),\\
& \varepsilon_0(R)=\varepsilon_d + g R, \quad
    U_0(R)=\tfrac{1}{2} m \omega^2 R^2,
\end{align}
where $\varepsilon_d$ is the LUMO energy at $R=0$, $g=\partial\varepsilon_0/\partial R$ is the (Condon) linear vibronic coupling, and the dependence on $x_M$ is omitted for notational simplicity.

In contact with an extended electronic system, vibrational relaxation can proceed via nonadiabatic excitation of electron--hole pairs (EHPs) \cite{persson1980vibrational, dou2020nonadiabatic}. Under the standard separation of timescales where the electronic subsystem equilibrates much faster than the nuclear motion \cite{spohn1980kinetic, dann2018time}, the vibrational energy loss rate is captured by an electronic friction kernel \cite{bohnen1975friction, persson1980vibrational, tully1995molecular,  huang2000vibrational, bunermann2015electron, ryabinkin2017mixed, dou2020nonadiabatic}. For a single coordinate $R$,
\begin{align}
    \gamma (R) &= - \pi \hbar \int_{-\infty}^{\infty}\! \mathrm{d}\epsilon\; \Xi(\epsilon;R)\, \frac{\partial f_T(\epsilon)}{\partial \epsilon}, \\
    \Xi(\epsilon;R) &= \mathrm{Tr}\!\left[ \frac{\partial \hat{H}(R)}{\partial R}\, \hat{P}(\epsilon;R)\, \frac{\partial \hat{H}(R)}{\partial R}\, \hat{P}(\epsilon;R) \right],
\end{align}
where the trace is over the single-particle electronic space and $\hat{P}(\epsilon;R)=\sum_\lambda \delta\!\big(\epsilon-\epsilon_\lambda(R)\big)\,|\lambda(R)\rangle\langle\lambda(R)|$ is the spectral projector. For the Hamiltonian in \eqref{eq:FA_Hm} with \eqref{Eq:epsilon0R}, assuming (i) only the adsorbate level $\varepsilon_0(R)$ depends on $R$ (no $R$-dependence in the couplings $T_{s,\alpha}$, i.e., with non-Condon effects neglected), and (ii) linear response about a fixed $R$, the friction reduces to
\begin{align} \label{eq:Friction}
   \gamma(R) &=  -\pi \hbar\, g^{2}\!\int_{-\infty}^{\infty}\! \mathrm{d}\epsilon \sum_{\lambda,\lambda'} 
   \big|\langle \lambda(R) \,|\, \varepsilon_0\rangle\big|^{2} \delta\!\big[\epsilon-\epsilon_{\lambda}(R)\big]\nonumber\\
   &\times \big|\langle \lambda'(R) \,|\, \varepsilon_0\rangle\big|^{2} \delta\!\big[\epsilon-\epsilon_{\lambda'}(R)\big] \frac{\partial f_T(\epsilon)}{\partial \epsilon}.
\end{align}

\tb{Equation~\eqref{eq:Friction} corresponds to the Head--Gordon--Tully electronic-friction tensor, i.e., the \emph{equilibrium, Markov} isolated system limit for nuclear motion coupled to a dense manifold of electronic excitations \cite{tully1995molecular}. We note that more general nonequilibrium open-system treatments can yield non-Markovian friction and additional force contributions beyond a simple Markovian friction tensor \cite{chen2018current}. Nonetheless, in the assumed equilibrium, isolated system limit with fast decoherence (induced by other molecular or substrate degrees of freedom), the HGT expression has been shown to be consistent with more general open-system formulations of electronic friction \cite{dou2017universality, chen2018current}.}

\tb{In our simulations the electronic environment is represented by a finite SSH chain, so the electronic spectrum is discrete at finite $N$. To obtain a continuous spectrum, and hence a well-defined Markovian friction, we broaden each discrete level by a Gaussian of width $\sigma$ following standard practice \cite{tully1995molecular,hellsing1984electronic}, 
replacing the Dirac delta functions by normalized Gaussians,
\begin{align} \label{eq:gaussian}
    \delta\!\big(\epsilon - \epsilon_\lambda\big) \;\to\; \frac{1}{\sigma\sqrt{2\pi}} \exp\!\left[-\frac{(\epsilon -\epsilon_\lambda)^2}{2\sigma^2}\right],
\end{align}
thereby yielding a smooth LDOS $P(\epsilon,R)$ and numerically stable friction coefficients \cite{askerka2016role}.}

\tb{In this work, we use an $N$-independent broadening parameter chosen on the order of the thermal smearing scale, $\sigma \sim k_B T$ \cite{aarons2016perspective}. Importantly, our conclusions do not depend on this choice: the friction coefficients are robust with respect to moderate changes in $\sigma$ and with respect to increasing the chain size $N$ (see \SI)}

%% file: results_and_discussion.tex
\par \noindent\textbf{Bulk and boundary chemisorption.}  \label{sec:res&dis}
In what follows, we examine the chemisorption of a single adsorbate on the SSH chain under various conditions. 

\par \tb{To place our parameter choices in the context of realistic 1D conjugated systems, we note that tight-binding descriptions of polyacetylene and polyacetylene-like chains \cite{mizes1991tight, meider1997density, wang2019surface, paloheimo1992density} are broadly consistent with nearest-neighbor $\pi$-electron hopping scales of a few eV and a moderate bond-alternation (dimerization) contrast. Representative values such as $v=2.5$~eV and $w=3.5$~eV \cite{mizes1991tight} imply a hopping ratio $w/v$ of order unity with appreciable bond alternation, consistent with the scale of parameters used in the present model.}

\tb{Because our central conclusions are governed by dimensionless ratios such as the dimerization ratio
\[
r \equiv \frac{w}{v},
\]
the dimensionless LUMO energy $\varepsilon_0/v$, and hopping $T/v$, the predicted midgap-state-enabled enhancement of hybridization and electronic friction is robust across the range of $r$ explored here. We therefore expect the reported trends to be transferable.}

\tb{Our analysis
covers the trivial (\(r<1\)), metallic (\(r\simeq 1\)), and topological (\(r>1\)) regimes. We emphasize our primary objective is not to model a specific material parameter set, but to map how the topological phase transition and the appearance of boundary-localized midgap states modulate charge hybridization and electronic friction for an adsorbate across a broad, physically motivated range of $r$. }

\par Throughout this subsection, the molecular geometry is fixed (\(R=0\)). The adsorbate is therefore characterized by its low-lying level energy \(\varepsilon_0\) and placement \(x_M\). \tb{We also note that all equilibrium observables reported below are converged with respect to the SSH chain length: increasing  the number of unit cells $N$ leaves the electron occupancy essentially unchanged so that the finite chain adopted here represents the system in thermodynamic limit (see \SI for verification).}

\par
\begin{center}
    \begin{figure}[hbt]
    \includegraphics[width=8cm]{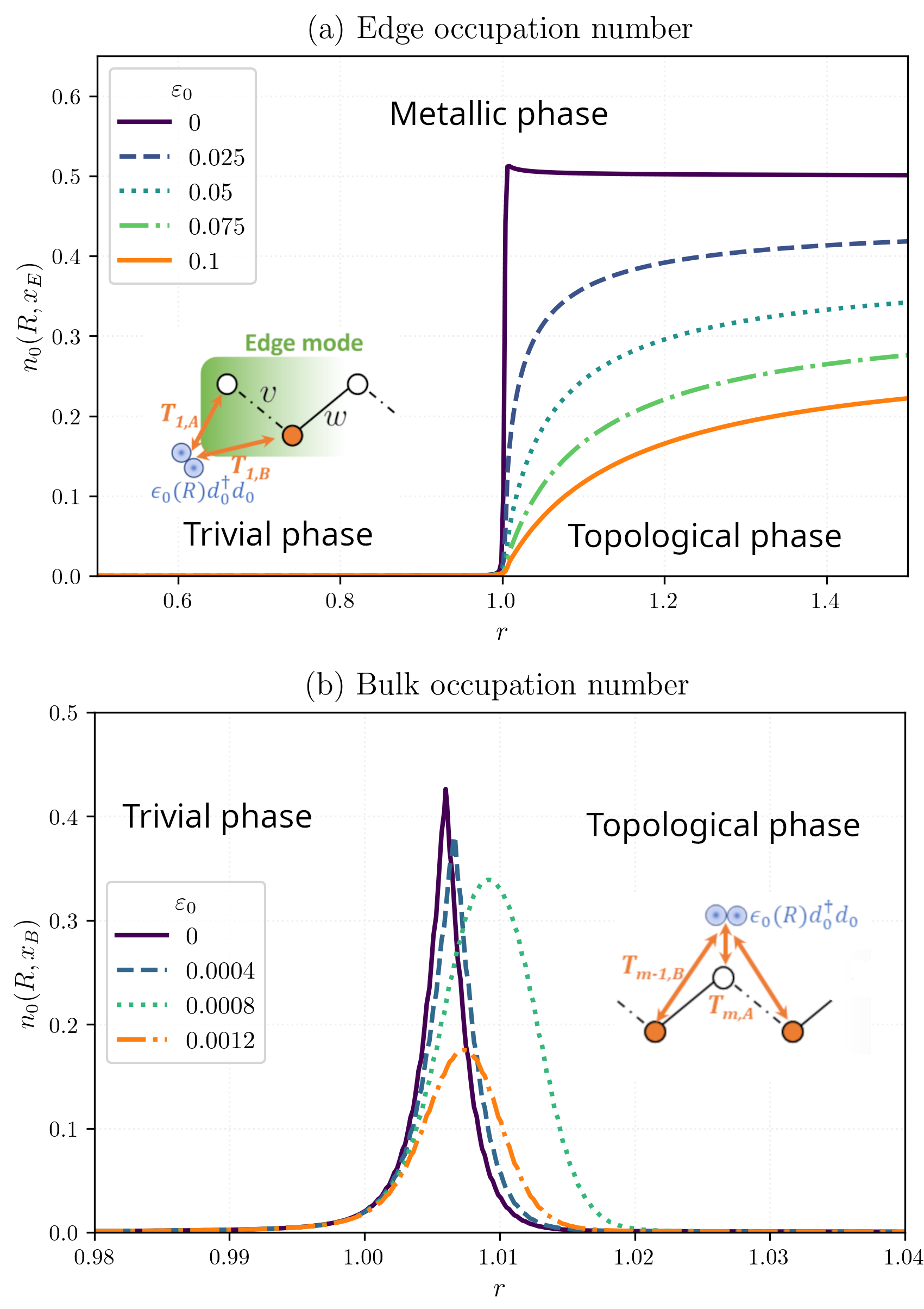}
    \caption{LUMO occupation near the (a) edge \((T_{1,A}=T_{1,B}/3=0.1)\) and (b) bulk center \((T_{N/2,A}=T_{N/2+1,B}/3=T_{N/2-1,B}/3=0.1)\) of the chain for trivial \((r<1)\), metallic \((r\approx 1)\), and topological \((r>1)\) phases. Parameters: \(N=800\), \(v=10\), \(\mu=0\).}
    \label{fig:Occupation_Number}
    \end{figure}
\end{center}

 Figures~\ref{fig:Occupation_Number}(a,b) report the LUMO occupation \(n_0(R,x_M)\) at 0~K and half filling (\(\mu=0^-\)) for adsorption near the edge (\(x_M=x_E\)) or at the bulk center (\(x_M=x_B\)). Figure~\ref{fig:Occupation_Number}(a) shows that the trivial insulator exhibits negligible electron sharing at the edge, whereas the topological phase displays strong hybridization with a sharp rise of \(n_0(R,x_E)\) upon entering \(r>1\). The enhancement is most pronounced when \(\varepsilon_0\) is nearly resonant with the boundary mode. Conversely, Fig.~\ref{fig:Occupation_Number}(b) shows that for adsorption in the metallic bulk, significant \(n_0(R,x_B)\) occurs only in a narrow neighborhood of the topological transition critical point (\(r\approx 1\)), where the density of occupied states near \(\varepsilon_0\) is largest. While the metallic phase maximizes bulk adsorption, boundary chemisorption in the topological phase remains the optimal overall scenario for hybridization under the considered conditions as \(\max n_0(R,x_B) < \max n_0(R,x_E)=0.5\). Notably, appreciable adsorbate-edge hybridization persists even when \(\varepsilon_0\gg 0\).

\begin{center}
    \begin{figure}[hbt]
    \includegraphics[width=8.5cm]{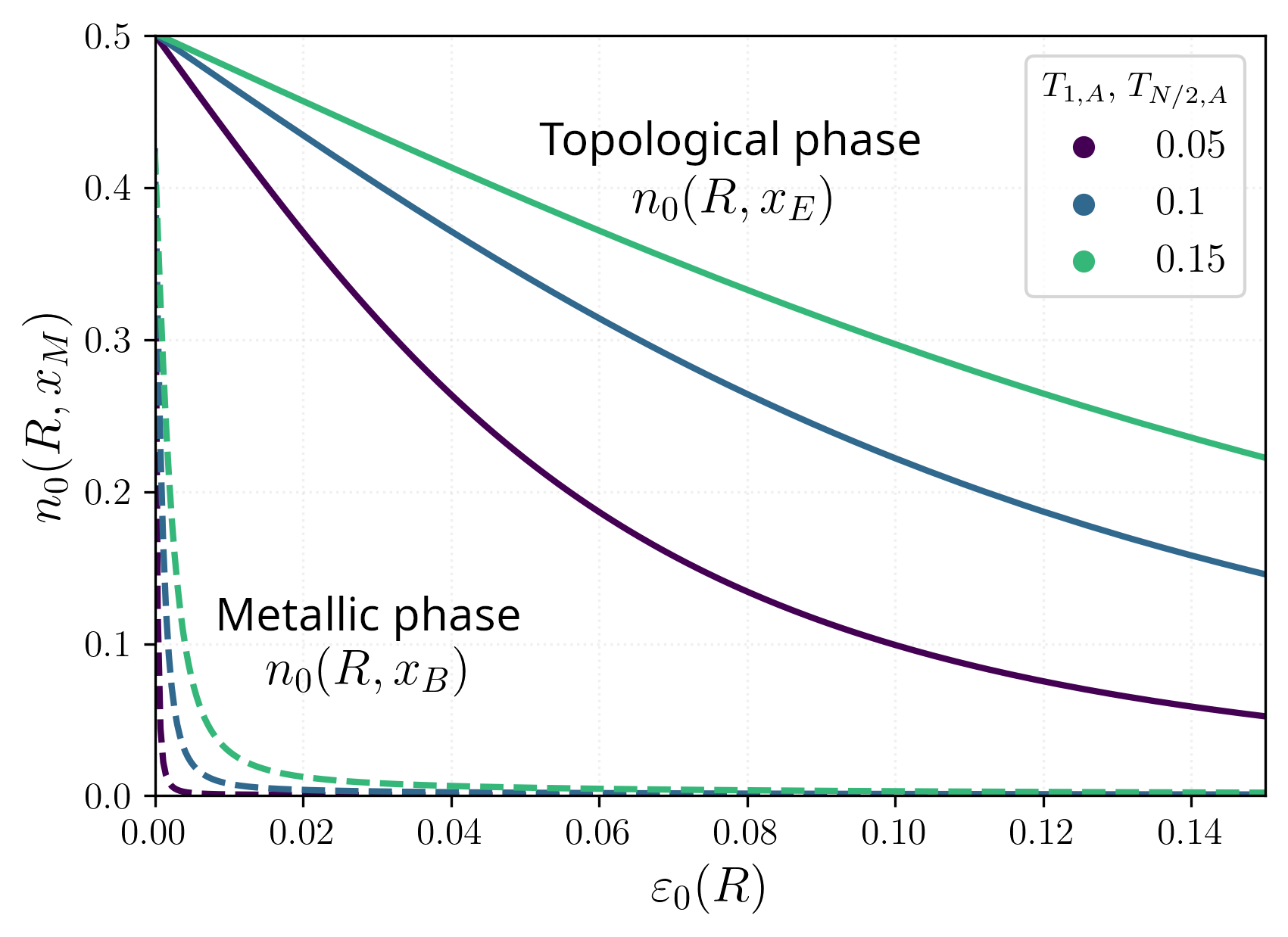}
    \caption{LUMO occupation vs.\ adsorbate level \(\varepsilon_0\): topological edge (\(w=15\), solid lines) compared with metallic bulk (colors denote \(r=1.00814\) [purple], \(1.00597\) [blue], \(1.00465\) [blue–green]) for several \(T_{m,A}\). Parameters: \(N=800\), \(v=10\), \(\mu=0\).}
    \label{fig:Occupation_Comparision}
    \end{figure}
\end{center}

The robustness of the topological enhancement is highlighted in Fig.~\ref{fig:Occupation_Comparision}, which directly compares edge adsorption in the topological phase to bulk adsorption in the metallic phase over varying \(\varepsilon_0\) and couplings \(T_{m,A}\). Across all cases examined, hybridization with metallic bulk states yields a lower \(n_0\) than coupling to the topological edge mode, despite the larger metallic DOS near \(\varepsilon_0\). Notably, edge \(n_0(R,x_E)\) remains substantial even when \(T_{1,A}\ll v,w\).

\textbf{Discussion.} 
In the topological phase and near resonance (\(\varepsilon_0\approx 0\)), the adsorbate couples predominantly to the nearest edge mode. For \(T_{1,A},T_{1,B}\ll v,w\) and \(w/v-1\gg 0\), the hybridized edge–LUMO doublet is well isolated from bulk continua, yielding the symmetric occupancy \(n_0(R,x_E)\approx 0.5\). In the metallic regime (\(r\simeq 1\)), \(\varepsilon_0\approx 0\) is near-degenerate with many occupied/unoccupied extended states and the resulting Fano line shape \cite{fano1961effects} in the local molecular DOS produces a lower maximal $n_0 \approx 0.4$. Off resonance, bulk adsorption admits the standard second-order estimate

\newcommand{\Tc}{\boldsymbol{T}_c}
\newcommand{\ukc}{\boldsymbol{u}_{\kappa,c}}
\begin{equation}\label{eq:Bulk_Non_Occ}
  n_0(R,x_B) \;\approx\; \sum_{ E_\kappa^{(0)}<\mu}
  \frac{\bigl|\Tc\!\cdot\!\ukc\bigr|^{2}}{\bigl(\varepsilon_0 - E_\kappa^{(0)}\bigr)^{2}},
\end{equation}
where $E_\kappa^{(0)}$ denote bare SSH eigenenergies, $c = N/2$ and $\Tc$ and $\ukc$ are 
\[
\Tc \equiv \begin{pmatrix} T_{c,A} \\[2pt] T_{c,B} \\[2pt] T_{c-1,B} \end{pmatrix},
\qquad
\ukc \equiv \begin{pmatrix} a_{c,\kappa} \\[2pt] b_{c,\kappa} \\[2pt] b_{c-1,\kappa} \end{pmatrix},
\]
and \(a_{j,\kappa}, b_{j,\kappa}\) are the amplitudes of eigenstate \(\kappa\) on sublattices \(A/B\) at cell \(j\). The sum runs over occupied states at \(T\!=\!0\) K. Away from criticality the occupied bulk states with non-negligible weight at the molecule are delocalized and detuned from \(\varepsilon_0\), suppressing \(n_0(R,x_B)\). Conversely, when \(r\approx 1\), the increased density of states and small denominators in \eqref{eq:Bulk_Non_Occ} enhance \(n_0\).

\par
\begin{center}
    \begin{figure}[b]
    \includegraphics[width=8.2cm]{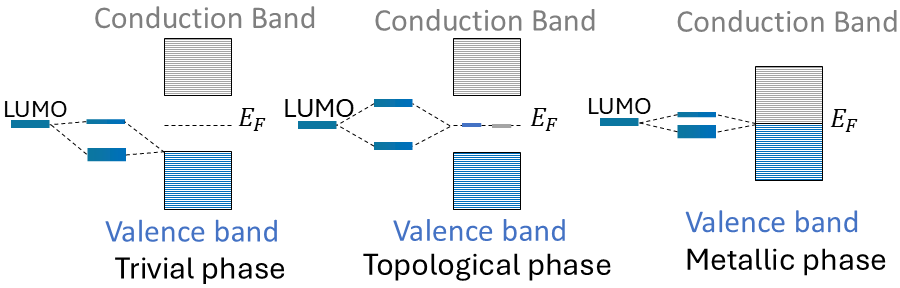}
    \caption{Schematic hybridization between the adsorbate LUMO and SSH states in the trivial (left), topological (center), and metallic (right) regimes.}
    \label{fig:mol_hybr}
    \end{figure}
\end{center}

 In summary (Fig. \ref{fig:mol_hybr}), 
we find that strong spatial localization of the topological midgap modes enables robust hybridization and charge donation into an edge-bound adsorbate. By contrast, despite its larger DOS, the metallic phase shows weaker hybridization and charge transfer into the adsorbate LUMO at a given local contact because extended states dilute the local overlap between the adsorbate low-lying orbital and the SSH chain. 

\noindent\textbf{Chemisorption on solitons and trivial defects}
In an ideal SSH chain, midgap ($E\!\approx\!0$) modes occur only at the edges of the topological phase $r > 1$. At finite temperature, bulk topological defects known as solitons (domain walls) \cite{su1979solitons, takayama1980continuum, heeger1988solitons} arise and bind additional exponentially localized midgap states in the interior. These localized states may also hybridize efficiently with nearby molecular orbitals, enabling charge donation to an adsorbate. In contrast, a metallic SSH chain (\(r=1\)) lacks localized midgap modes, but has greater density of states of near-resonant levels with the adsorbate, and in the presence of trivial defects (e.g., impurities) may also have local resonances that could hybridize effectively with adsorbates. We therefore compare multi-adsorbate chemisorption on the following substrates: (i) a topological chain with multiple solitons (soliton ensemble), (ii) an ordered metallic chain, and (iii) a disordered metallic chain with trivial defects.

\par
\begin{figure}[t]
  \centering
  \includegraphics[width=8cm]{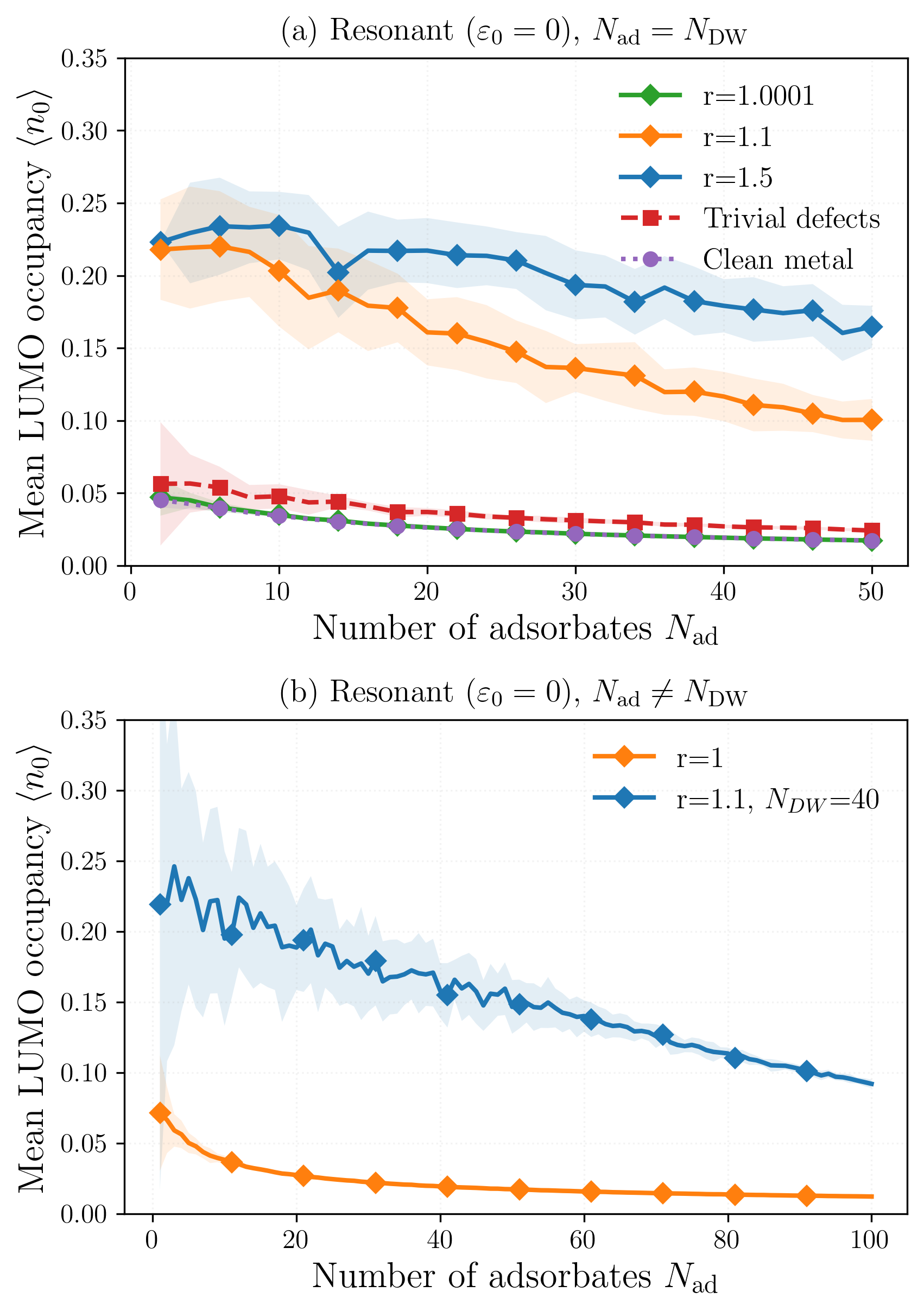}
  \caption{Resonant case \(\varepsilon_0=0\). 
  (a) \(\expval{n_0}\) vs.\ number of adsorbates (defects) for soliton ensembles at three dimerizations (\(r=1.5,1.1,1.0001\)), compared to (i) metal (\(r=1\) with the same number of on-site defects and (ii) disorder-free metal (\(r=1\), no defects). Shaded regions represent fluctuations over realizations of the disordered systems, and we have set the number of adsorbates to be equal the number of domain walls \( N_{ad} = N_{DW} \).
  (b) \(\expval{n_0}\) vs.\ \(N_{\text{ad}}\) for a fixed set of 40 domain walls (topological) or none (zero-disorder metal), and adsorbates placed at random sites. Parameters: \(\xi=7\), \(N=3000\), \(v=10\), \(T_m=0.1\), and 25 realizations.}
  \label{fig:in-soliton}
\end{figure}

 We report the average LUMO occupancy \(\langle n_0\rangle\) [Eq.~\eqref{eq:average_n0}] for \(N_{\text{ad}}\) adsorbates on a chain of \(N\) unit cells at 0~K and half filling. Domain walls have width \(\xi\in\{5,7,10\}\) (corresponding to the localization length of their midgap states, see Eq.~\eqref{eq:phinm0}). In the soliton ensemble simulations reported in Fig. \ref{fig:in-soliton}a), each adsorbate sits at a soliton center and \(N_{\text{ad}}\) equals the number of domain walls \( N_{DW} \) (including the two edges). In the metallic phase with trivial defects, we set \(r=1\) and introduce a dilute ensemble of defects by drawing small near-zero on-site potentials from a Gaussian distribution, applied to a random subset of sites to mimic weak non-topological disorder. Adsorbates are then placed at defect locations drawn from a uniform distribution over bulk sites. Finally, in the zero disroder metal scenario, $r=1$ and the adsorbates are placed at uniformly random bulk sites. Unless otherwise specified, these simulations employed \(N=3000\), \(v=10\), \(T_m=0.1\), and averages over 25 realizations with fixed electron number.

\begin{figure}[t]
  \centering
  \includegraphics[width=8cm]{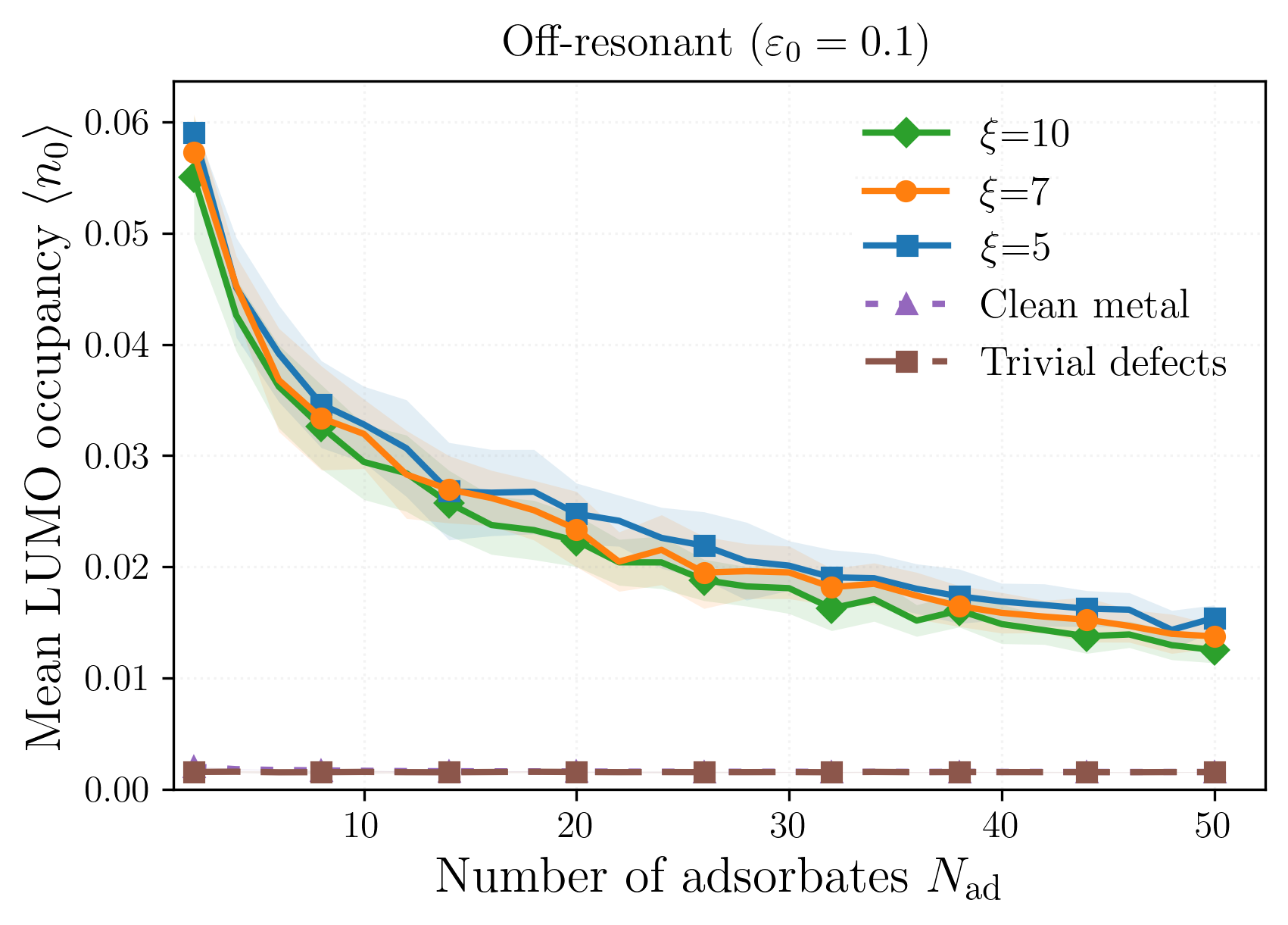}
  \caption{Off-resonant case \(\varepsilon_0=0.1\): \(\braket{n_0}\) vs.\ \(N_{\text{ad}}\) and defect content. Parameters: \(N=3000\), \(v=10\), \(T_m=0.1\).}
  \label{fig:off-soliton}
\end{figure}

\par Across all conditions, chains with domain walls yield systematically larger \(\expval{n_0}\) than metallic chains with trivial defects (Fig.~\ref{fig:in-soliton}a). The enhancement tracks soliton localization: larger \(r\) (stronger dimerization) narrows midgap wavefunctions and increases adsorbate–chain overlap \cite{su1979solitons}. Near criticality (\(r=1.0001\)) midgap states broaden and charge transfer is reduced. Fluctuations reflect variability from soliton–antisoliton spacing and boundary proximity.

The near plateau $\expval{n_0} \approx 0.25$ at $r=1.1$  (Fig. \ref{fig:in-soliton}a) can be explained as follows.
At half filling and with particle–hole symmetry, each isolated domain wall binds one midgap state at zero energy which is half filled on average. Given that any adsorbate positioned in the neighborhood of a particular soliton can reach at most \(n_0 \approx 0.5\) when $\varepsilon_0 \geq 0$ (Fig. \ref{fig:Occupation_Comparision}), averaging over all adsorbates gives a low-coverage plateau of \( \expval{n_0} \approx (0.5)\times(0.5)=0.25\). As the number of domain walls increases, domain-wall interactions lead to the observed decay in \(\expval{n_0}\). Metals (\( r = 1 \)) with trivial defects behave similarly to clean metals, i.e., both give much lower $\expval{n_0}$ relative to the system with topological defects, thus indicating that non-topological near-zero resonances do not generically replicate midgap-assisted donation and instead hybridize  significantly with metal extended states rather than to the adsorbate orbital.

Figure~\ref{fig:in-soliton}b fixes forty solitons but randomizes adsorbate positions, hence, the adsorbates may or may not land on a domain wall. At low coverage the same plateau \(\expval{n_0}\!\approx\!0.25\) emerges, but with large variance: realizations in which adsorbates overlap midgap states yield \(n_0\!\sim\!0.5\), whereas those on bulk sites yield \(n_0\!\sim\!0\). This two-population mixture (soliton-bound vs. bulk-bound) produces both the \(0.25\) mean and the broad fluctuations (shaded blue region). In contrast, ordered metallic chains give substantially lower mean LUMO occupancy at all coverages, underscoring the chemisorption advantage conferred by topological midgap states.

\par Off resonance (\(\varepsilon_0>0\); Fig.~\ref{fig:off-soliton}) the topological advantage persists but is reduced. The dependence on \(\xi\) is weak-to-moderate: narrower domain walls (smaller \(\xi\)) produce more localized midgap states and slightly larger \(\expval{n_0}\) due to enhanced wavefunction overlap. In contrast, the trivial-defect metal behaves similarly to the ordered metal, indicating that non-topological near-zero local resonances do not generically replicate the midgap-assisted donation of domain walls.

\par
\begin{figure}[t]
  \centering
  \includegraphics[width=8.5cm]{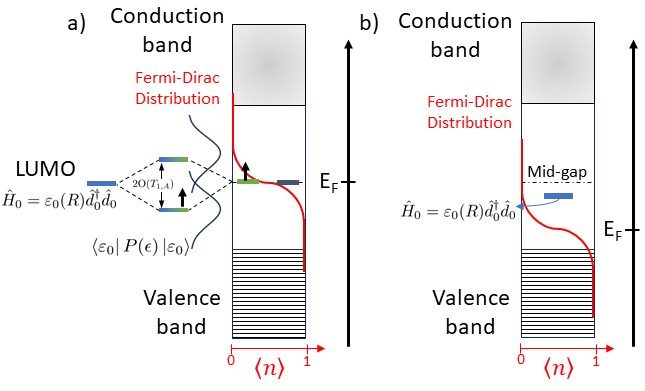}
  \caption{(a) Topological edge: hybridization between the localized edge state and the adsorbate level produces a PDOS (blue solid lines) doublet \(\langle \varepsilon_0|P(\epsilon,R)|\varepsilon_0\rangle\) split by \(\sim 2\,\mathcal{O}(T_{1,A})\). Coupling to the distant edge is negligible. (b) Trivial insulating bulk: with half filling, \(E_F\) sits near midgap; the adsorbate level is far from the band edges, yielding vanishing \(\gamma(R)\).}
  \label{fig:Friction_FD1}
\end{figure}

 \noindent\textbf{Topological phase transition signature on adsorbate electronic friction.} \label{ssec:results_friction} 
Using the adiabatic orbitals from diagonalizing Eq.~\eqref{eq:FA_Hm} with \(\varepsilon_0(R)=\epsilon_d+R\,g\sqrt{m\omega/\hbar}\), we construct the local molecular density of states \(P(\epsilon,R)\) and evaluate the electronic friction \(\gamma(R)\) via Eq.~\eqref{eq:Friction} for the metallic (\(r=1\)), topological (\(r>1\)), and trivial (\(r<1\)) phases at temperature \(1/\beta=0.015\). As explained above, \(P(\epsilon,R)\) is obtained by Gaussian broadening of each adiabatic level \(\epsilon_\lambda(R)\) with width \(\sigma=0.0225\). While \(\sigma\) is not uniquely defined \cite{maurer2016ab}, this choice yields smooth non-oscillatory \(P(\epsilon,R)\) and \(\gamma(R)\) sufficient for qualitative trends across phases \tb{(see \SI for a sensitivity analysis)}. 
For a dilute impurity concentration and fixed electron number in the polyacetylene chain, we treat the system effectively as intrinsic \cite{ashcroft2022solid}. We modeled the interacting adsorbate as a hole-like impurity and set the Fermi level $\mu$ in the HOMO–LUMO midgap at half filling (spinless electrons, for \(N\) unit cells). Because hybridization shifts the spectrum, $\mu$ is tracked self-consistently as a function of \(R\); see Fig.~\ref{fig:Friction_FD1}b.

\begin{figure}[t]
  \centering
  \includegraphics[width=8.5cm]{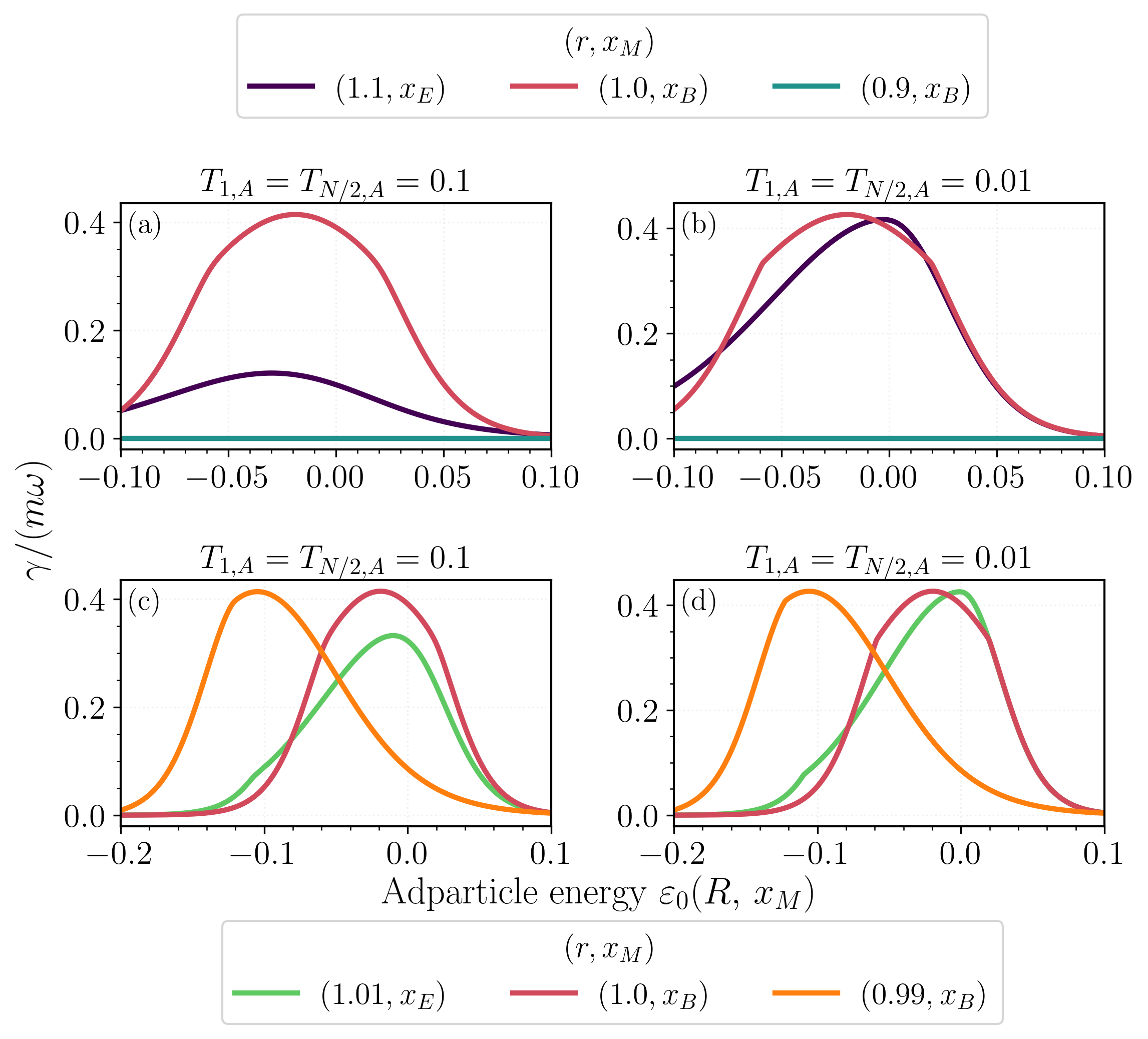}
  \caption{Electronic friction \(\gamma(R)\) [Eq.~\eqref{eq:Friction}] across phases.
  (a) Topological edge (\(x_E\); purple, \(r=1.1\)), metallic bulk (\(x_B\); magenta, \(r=1\)), and trivial bulk (\(x_B\); teal, \(r=0.9\)) at \(T_{1,A}=T_{N/2,A}=0.1\).
  (b) Same as (a) with weaker coupling \(T_{1,A}=T_{N/2,A}=0.01\).
  (c,d) Near criticality: topological edge (\(x_E\); green, \(r=1.01\)) and trivial bulk (\(x_B\); orange, \(r=0.99\)) vs.\ \(\varepsilon_0(R)\) at \(T_{1,A}=0.1\) (c) and \(T_{1,A}=0.01\) (d).
  Parameters: \(v=10\), \(\sigma=0.0225\), \(\varepsilon_d=0.15\), \(g=0.02\), \(\beta=1/0.015\), \(N=800\), with \(T_{N/2-1,B}=T_{N/2,B}=T_{N/2,A}/3\).}
  \label{fig:Friction}
\end{figure}

Figure~\ref{fig:Friction}a shows \(\gamma(R)\) vs.\ \(\varepsilon_0(R)\) for representative placements and dimerization $r = v/w$. The metallic phase exhibits the largest friction, consistent with a high LDOS at \(\mu\) and abundant low-energy electron–hole excitations. In the topological phase at the edge, the localized boundary mode enhances friction relative to a gapped bulk but remains below the metallic value. The reduction arises from the hybridization-induced splitting between the edge mode and the LUMO: a two-level avoided crossing produces a PDOS doublet \(\langle \varepsilon_0|P(\epsilon,R)|\varepsilon_0\rangle\) separated by \(\Delta E\sim 2\mathcal{O}(T_{1,A})\), which depresses the LDOS at \(\mu\) and suppresses electron-hole pair generation (Fig.~\ref{fig:Friction_FD1}a) near the Fermi level.

\par At smaller adsorbate–edge coupling \(T_{1,A}=0.01\) [Fig.~\ref{fig:Friction}b], the splitting \(\Delta E\) diminishes, partially restoring LDOS at \(E_F\) and increasing \(\gamma(R)\) relative to the stronger-coupling case. Note this enhancement occurs as one reduces a strong coupling. For sufficiently weak coupling, $\gamma$ ultimately decreases with $T_{1,A}$ as the adsorbate decouples (see \SI for further discussion).

\par For a sizeable SSH gap \(|w-v|=1\) (e.g., \(v=10\), \(w=9\)), thermal carriers are exponentially suppressed and the adsorbate level lies far from the valence-conduction continua [Fig.~\ref{fig:Friction_FD1}b]. With \(\mu\) pinned near midgap, \(\gamma(R)\) is negligible in the trivial phase, consistent with the absence of low-energy electron-hole channels. Likewise, in the topological phase but \emph{in the bulk} (away from edges), \(\gamma(R)\) is similarly suppressed; we therefore omit those curves from Fig.~\ref{fig:Friction} for clarity. In such gapped settings, vibrational energy relaxation is likely dominated by phonon-mediated channels in the insulating substrate \cite{sjakste2018energy,giustino2017electron}.

Figures~\ref{fig:Friction}c,d examine \(\gamma(R)\) as \(\varepsilon_0(R)\) sweeps across band edges close to criticality. In the topological phase, the friction grows when \(\varepsilon_0(R)\) approaches the band edges because both conduction and valence states contribute to \(P(\epsilon,R)\) near \(E_F\). In the trivial phase, the friction peak shifts toward the valence band edge. The maximum occurs when the adsorbate aligns with the bulk HOMO, \(\varepsilon_0(R)\approx w-v\). For \(v=10\) and \(r=0.99\) (i.e., \(w=9.9\)), this gives \(\varepsilon_0(R)\approx -0.1\). Reduced \(T_{1,A}\) narrows the avoided crossing even near criticality, mirroring the trend in Fig.~\ref{fig:Friction}b.

%% file: conclusions.tex
\label{sec:conclusions}
\par We studied chemisorption of adsorbate species with an empty frontier level coupled to a polyacetylene SSH chain across its trivial (\(r<1\)), metallic (\(r=1\)), and topological (\(r>1\)) regimes. We extensively analyzed (i) electron donation into adsorbate LUMO and (ii) nonadiabatic electronic friction experienced by an adsorbate vibrational coordinate as a function of adsorbate position along the chain and its internal geometry. We further contrasted topological domain walls (solitons) with non-topological, trivial defects in the metallic phase.

\par We find a robust advantage for electron donation in the topological phase relative to both the metallic and trivial phases. This is explained by the fact that exponentially localized midgap states at edges (and at solitons) have large wavefunction amplitude at the adsorption site, which enhances adsorbate–substrate hybridization at resonance. By contrast, although the metallic phase has a larger total density of states near the Fermi level, its extended bulk states are spatially dilute at any one site, weakening local hybridization and reducing charge transfer. Importantly, merely generating localized states is \emph{necessary but not sufficient}: the donation depends also on resonance alignment, spatial overlap, and occupancy. Topological midgap modes satisfy these criteria in a robust, symmetry-protected manner, whereas trivial near-zero resonances in a metal generally do not. Consistent with this picture, solitons in the topological phase substantially outperform trivial defects in a metal. Near criticality (weak dimerization regime $r \approx 1$) the midgap states broaden and the advantage correspondingly diminishes.\\

\par Adsorbate electronic friction provides a complementary dynamical signature of the electronic phase structure. It is largest in the metallic regime, where abundant low-energy electron–hole excitations exist at the Fermi level. In the topological phase at an edge, friction is enhanced relative to a gapped bulk but remains below the metallic value because edge–LUMO hybridization produces an avoided crossing that splits the molecular projected density of states and depresses the local density of states precisely around \(\mu\). In the trivial phase, and in the bulk of the topological phase away from edges, the electronic gap suppresses friction. These trends persist across adsorbate–substrate coupling strengths, with weaker coupling reducing the hybridization splitting and partially restoring the LDOS at \(\mu\).

\par These results suggest that low-dimensional topological substrates enhance adsorbate hybridization by exploiting localized midgap states at edges and domain walls for charge donation, and they promote adsorbate vibrational energy dissipation via enhanced electronic friction. Extensions to higher dimensions are promising: topological materials with localized boundary modes and appreciable boundary LDOS near $\mu$ (e.g., three-dimensional topological insulators or topological semimetal surfaces with robust surface states) are expected to further strengthen both electron donation and adsorbate friction, subject to the same resonance and overlap constraints discussed in this study. Finally, strategically patterning domain walls or stabilizing edge-rich morphologies offers a practical route to translate topological midgap physics into tunable knobs for adsorption energetics and vibrational relaxation in catalysis and surface chemistry. 

\par \tb{Finally, our choice of a spinless, weakly correlated SSH model ensures our modeling captures symmetry-protected midgap orbitals at edges and domain walls and their hybridization with an adsorbate. In more complex topological materials of interest for catalysis, a detailed treatment of spin degrees of freedom, strong spin--orbit coupling, and  strong electron--electron correlations may reshape the low-energy spectrum, change the relevant symmetry class, and open additional relaxation channels. Extending the present framework to spinful tight-binding models (including SOC) and to beyond-mean-field correlated descriptions (e.g., SSH--Hubbard-type models) is an important direction for quantifying the robustness of the trends identified here and for exploring regimes where electronic correlations qualitatively alter charge transfer and dissipation at molecule--surface interfaces.}